\documentclass[fleqn, onecolumn, usenatbib, useAMS]{mnras}
\usepackage{graphicx}	% Including figure files
\usepackage{float}

\usepackage{amsmath}	% Advanced maths
\usepackage{subcaption}
\usepackage{amssymb}	% Extra maths symbols
\usepackage{multicol}        % Multi-column entries in tables

\usepackage{bm}		% Bold maths symbols, including upright Greek
\usepackage{pdflscape}	% Landscape pages
\usepackage{hyperref}
\usepackage{blindtext}
\def\bea{\begin{eqnarray}}
\def\eea{\end{eqnarray}}
\def\ba{\begin{array}}
	\def\ea{\end{array}}

\def\beq{\begin{equation}}
\def\eeq{\end{equation}}

\newcommand{\eq}{Eq.\eqref}
\newcommand{\fig}{Fig.\ref}

\usepackage[T1]{fontenc}
\usepackage{ae,aecompl}

\usepackage{times}
\title[SNLS ]{\textsc{Estimating weak lensing convergence  correlation of Type-Ia supernovae from  5-year SNLS data by internal error estimate technique}}
\author[Mitra et al.]{Ayan Mitra$^{1}$\thanks{Contact e-mail: ayan.mitra@nu.edu.kz},
Arnab Paul$^{2}$\thanks{Contact e-mail: arnabpaul9292@gmail.com}, 	Barun Pal$^{3}$\thanks{Contact e-mail: terminatorbarun@gmail.com, Corresponding Author},
	Supratik Pal$^2$\thanks{Contact e-mail: supratik@isical.ac.in}\\
	$^{1}$Energetic Cosmos Laboratory, Nazarbayev University, Nur-Sultan, Kazakhstan -- 010000\\
		$^{2}$Physics and Applied Mathematics Unit, Indian Statistical Institute, Kolkata -- 700108, India\\
	$^{3}$ Department of Mathematics, Netaji Nagar College for Women, Kolkata -- 700092, India}
\begin{document}
\label{firstpage}
\pagerange{\pageref{firstpage}--\pageref{lastpage}}
\maketitle

\begin{abstract}
We report non-zero  weak lensing convergence correlation signal of Type-Ia supernovae from 5-year Supernovae Legacy Survey data. For our analysis we  utilize 296 supernovae magnification data from 5-year SNLS in the weak lensing limit. The data we use consists  of measurements from four different patches, each covering 1 square degree of the sky, merged together. We demonstrate that it is possible to have a very good estimate of the two point correlation function from this data using internal error estimate technique. In order to have a good estimate of the corresponding  covariance matrix we apply bootstrap spatial re-sampling technique where we reshuffle the original data consisting of 296 data points 100-10000 times and compare the results with that obtained from original data points. We show that this technique helps us arrive at a reliable conclusion on weak lensing convergence even though the original dataset comprises of a small number of data points. This also allows us to compute the corresponding covariance matrix with great accuracy.   
\end{abstract}


\section{Introduction}
Nearly two decades have passed since the discovery of accelerating universe from the measurement of luminosity distances of Type Ia Supernovae (SNe Ia) \cite{riess1998, Perlmutter1999}. Since then the surveys of  SNe Ia have   become an essential ingredient of modern cosmology. Till date SNe Ia are the most powerful probe for  the recent expansion history of our universe \cite{knop2003, barris2004, riess}.  The luminosity distance as measured from the  SNe Ia is a direct probe for the Hubble parameter which in turn  constrains the dark energy density and equation of state. 
Supernova Ia have  homogeneous intrinsic brightness, since, upon standardization, the dispersion in intrinsic brightness distribution turns out to be  $\le16\%$, 
thereby making them very good standard candle candidates. They can be used as an efficient cosmological probe of the expansion history of the universe. Using these observations \cite{riess}, \cite{perl} demonstrated the recent accelerated expansion of the Universe usually attributed to the cosmic constituent called dark energy. 

In supernova  cosmology, gravitational lensing effects the brightness distribution of the high redshift SNe Ia \cite{hol}. This can be seen as a source of uncertainty in the brightness distribution because of the additional scatter. This, however, has a sub-dominant  effect in the cosmological parameter estimation since  SNe magnification averages out due to flux conservation. 
In spite of that, in the following years gravitational lensing of the SNe Ia has emerged as an additional probe for the growth of cosmic structures. 
The study of gravitational lensing of SNe Ia (also standard candles) is useful as they help us infer about the mass distribution of the foreground line of sight. This can be materialized by analyzing their lensing magnification \cite{kron}, \cite{jonsson}, \cite{wil}, \cite{menard}, \cite{cooray2006}. SNe Ia light curves are calibrated using colour and stretch factor corrections \cite{phil}. These can be compared to the average standard candle light curves and conclusion can be drawn depending on their fainter or brighter nature than the average.

Simply put, gravitational lensing is the bending of light rays in the presence of a massive gravitational object in its vicinity \cite{lensing4, lensing5}. Lensing of SNe Ia concerns the distribution of the Hubble residuals, which provides information about the gravitational perturbations along the line of sight due to the introduction of non-Gaussian scatter into the SNe Ia Hubble diagram \cite{scovacricchi2016}.  SNe Ia lensing is an efficient tool for studying mass distribution of the large scale structures in the Universe as well as of galaxies and cluster halos \cite{lensing1,lensing2, lensing3} without depending on the nature of the matter.    There are two main observables in  lensing: shear $\gamma$ and convergence $\kappa$. Using shear one can do image analysis of the lensing sources and do foreground mass reconstruction by comparing the distortion with the un-lensed image sources. The other parameter,  convergence, which is  the dimensionless surface mass density, imparts an isotropic focusing effect on the lensing source (in contrast to shear  \cite{cooray2006} which produces an asymmetric distortion (anisotropy), thus lending a round source into an elliptic one). In lensing, magnification is a phenomenon responsible for the change in the apparent magnitude of the source. Mathematically it is given as 
\bea
\mu=\left|\det A\right|^{-1},
\eea
where $A$ is the distortion matrix, commonly known as the Jacobian matrix that takes into account the local properties of mass distribution.  In terms of
convergence and shear,  its determinant is given by 
\bea
\left|\det A\right| = (1-\kappa)^2-|\gamma|^2
\eea 
Since, in the weak lensing regime both $\kappa\leq1$ and $\gamma\leq1$, so essentially magnification reduces to 
\bea
\mu\simeq(1+2\kappa).
\label{eq3}
\eea
Although the higher order terms may play important role \cite{menard2003}, still the above approximation serves pretty well for extracting useful information from the convergence.

In this work our primary intention is to extract out meaningful information from the two point statistics done on the SNLS 5-year spectroscopic SNe Ia sample lensing data. 
More precisely, we will determine the two point correlation function of the lensing convergence from the magnification of SNe Ia using SNLS 5-year lensing data. The detection of cosmic magnification has come by cross-correlating fluctuations in background source count with the lower redshift foreground line of sight galaxies \cite{lensing5}. The first reliable detection of cosmic magnification came from the Sloan  Digital Sky Survey more than a decade ago \cite{scranton2005}. The angular auto-correlation of the magnification provides a direct measure of the lensing power spectrum containing the information about background cosmology and the growth of structure \cite{cooray2006, scovacricchi2016}. 

Our approach in this paper is  different from the usual analysis as it will be revealed subsequently. 
%Though  their analysis mostly deals with two point statistics of cosmic shear, we have employed to some extent, a  similar method to determine the two point statistics of the lensing convergence field, which is our point of consideration. 
The methodology used in this article focuses on finding the correlation function of the lensing convergence ($\kappa$) for the SNe Ia directly from SNLS 5 lensing data.   Keeping in mind that we only have 296 data points we used internal error estimate technique, viz. bootstrap resampling which allows to compute the covariance matrix very accurately to some extent.  The dataset we choose in order to materialize this
consists of the real data points for 5-year SNLS data as discussed in the previous section. Using this we shall show how one can extract out meaningful information for two point lensing correlation function for weak lensing convergence of SNIa.  This will also help us in comparing the results with the analysis done
in the rest of the article using spatial resampling and see if we can improve on the results done with the original data points. 
We reshuffle the original data consisting of 296 data points from 100-10000 times and compare the results with that obtained from original data points. We show that this technique helps us arrive at a reliable conclusion on the estimation of the weak lensing convergence estimator even though the original dataset comprises of a small number of data points. This also allows us to check for the consistency in the computation of the corresponding covariance matrix and  forecast on the  greater accuracy one can expect with the bigger sample size, expected in future datasets.   
However, in order to keep life simple,  we did not exploit the redshift information contained in the broad $z$ range of the SNe sample used here.

The outline of this paper is as follows : In Section \ref{data} the data is presented, the following Section \ref{methodology} we analyze the data without spatial resampling. Then in Section \ref{bootstrap} we apply spatial resampling technique to examine the data and compute the corresponding covariance matrix. In Section \ref{com} we have compared the results with and without spatial resampling and finally conclude the work in Section \ref{conclusion}.

\section{The Data}\label{data}
\begin{figure}%[H]
	\begin{center}
		\includegraphics[scale=0.55]{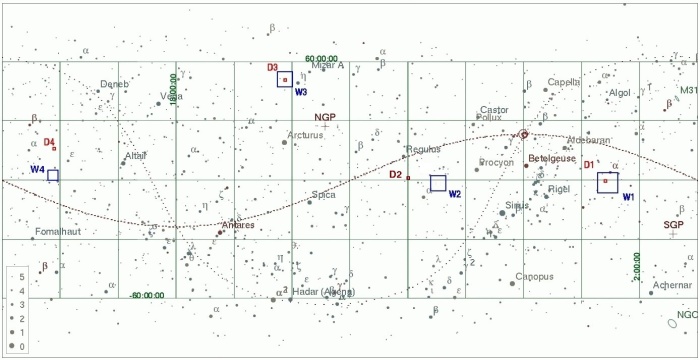}
	\end{center}
	\caption{Plot of the Deep (and the Wide field) survey locations on the full sky map. Image : \href{http://www.cfht.hawaii.edu/Science/CFHLS/cfhtlsdeepwidefields.html}{CFHTLS} website.}
	\label{f1}
\end{figure}

\begin{figure}%[H]
	\begin{center}
		\includegraphics[width=15cm, height=11.cm]{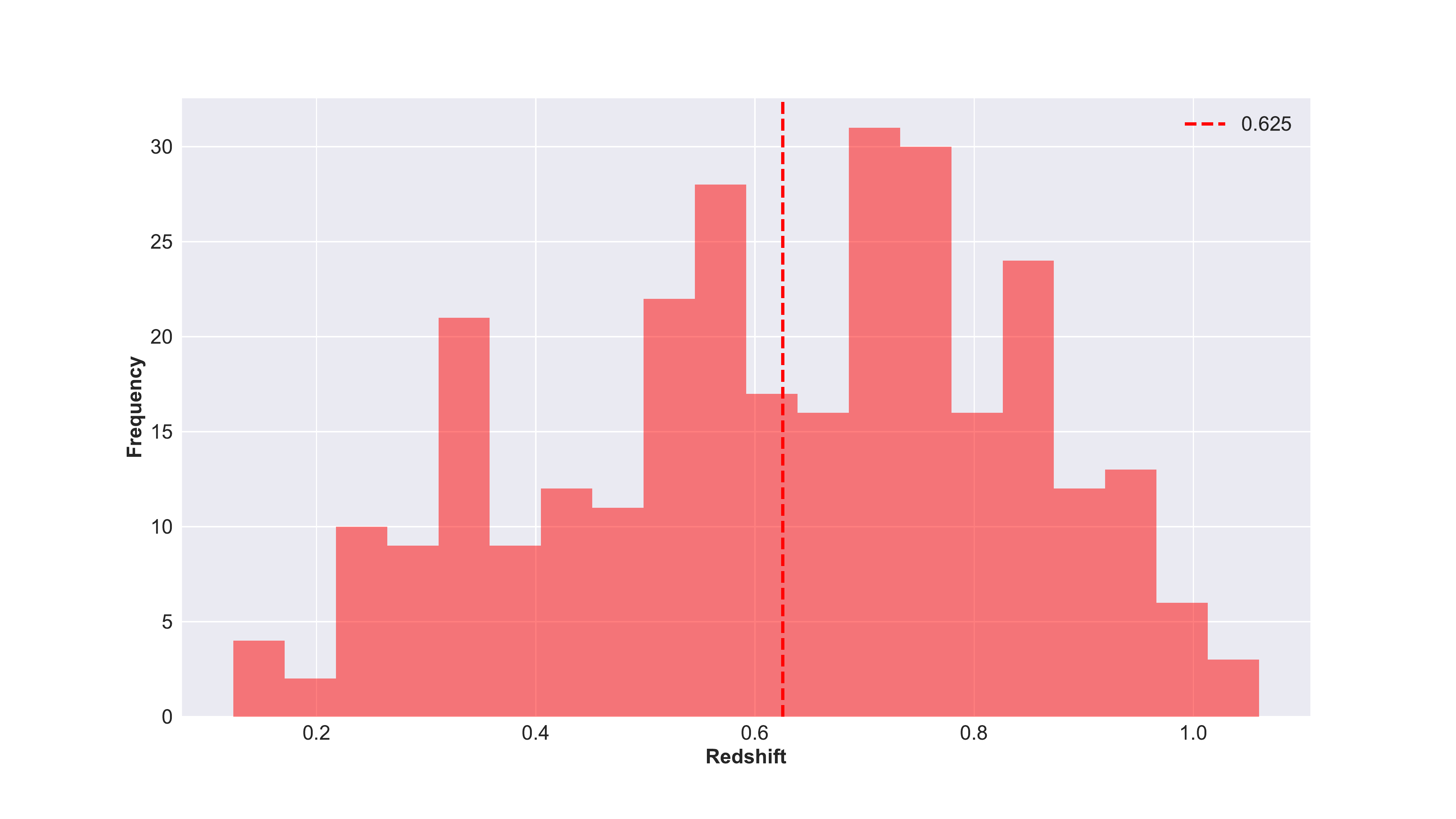}
	\end{center}
	\caption{Redshift distribution of the data. The dashed vertical red line shows the mean redshift ($z=0.625$) of the SNe Ia sample. The vertical axis shows the number of SNe Ia corresponding to a particular redshift.}
	\label{f2}
\end{figure}

The data used here consists of the SNe Ia sample provided in the SNLS 5-year lensing analysis. The detailed description of the data is available in Ref.\cite{mitra2016}. The SNLS used   CFHTLS Deep survey SNe Ia data which was drawn between $2003-2008$ with $450$ nights of observations. The CFHTLS used the MegaCam camera set up on the $3.6$ m Canada-France Hawaii telescope \footnote{\url{http://www.cfht.hawaii.edu/Instruments/Imaging/Megacam/}}.  The data was divided in four individual fields with $\sim1$ square-degree field of view in each of them.   The total number of SNe Ia used for our analysis is $296$.  The four Deep fields were separate patches of observations in the sky in four different locations taken in different seasons (\fig{f1}). 
Such narrow pencil beam like survey design enables to observe very distant supernovae, $z\sim1.2$. Table  \ref{tab1} gives the details of the four Deep fields. In \fig{f2} the redshift distribution of the lensing sample used for the analysis is presented. The dashed vertical line shows the mean redshift distribution of the SNe Ia sample $z=0.625$.  For the correlation computation, we carried out the analysis with the four individual $D$ field's  SNe Ia separately, which enabled us to check the correlations between the SNe Ia measured within the same patch of the sky in a single season. In each D-field we have computed the two point functions for each pair of SNe Ia, giving $n\times(n-1)/2$ number of points for $n$ samples in that field.  These points from the four individual fields  in the final analysis were merged together to obtain the final result. We used the convergence ($\kappa$)  for each SNe Ia obtained from their corresponding lensing magnification ($\mu$) expression given in \eq{eq3} for this analysis.  The convergence $\kappa$ that we used are retraced from the $\mu$ provided for each SNe Ia in \cite{mitra2016}. We made an approximation of extracting a single $\kappa$ for each SNe Ia data corresponding to each $\mu$. Originally, $\mu$ is computed by adding the effects of individual line of sight lenses given in the form of $\kappa_i$. However for our investigation purpose we compute a single mean $\kappa$ per SNe Ia using \eq{eq3}. We describe it as the mean projected inhomogeneity in matter distribution along the line of sight of each SNe Ia. The magnification values $\mu$ used from \cite{mitra2016} are computed using weak lensing approximation for each SNe Ia, in a circular aperture of $60''$ centered on the SNe within which the effect of lensing from galaxies are considered. Specifically, this is done, by computing the lensing (isotropic magnification effect in this case, parametrized by the convergence $\kappa$) effect produced by the halos of the individual line of sight galaxies. These halo models (eg. NFW or SIS) based on the galaxies properties and  their respective parameters (eg. velocity dispersion $\sigma$, rotational velocity etc.)  produce the corresponding lensing convergence (and magnification $\mu$, \eq{eq3}) for the SN Ia, hence acting as a lens to the source candidate. The cumulative effect of all these individual galaxies (and their haloes) acting as line of sight lenses, produce the net magnification. Hence there is a direct relation (and effect) between the foreground galactic structures (i.e. lenses) and the net effect of the SNe Ia magnification magnitude. 
Similar technique for SNe Ia lensing computation were presented before in Refs. \cite{ jonsson,kron} using smaller sample size and lesser detection accuracy. 
\begin{table}
	\centering
	\begin{tabular}{||c c c c c c c c c c||} 
		\hline
		\tiny Field & \tiny RA & \tiny Dec & \tiny E(B-V) &\tiny Other Observations & \tiny u & \tiny g & \tiny r & \tiny i & \tiny z\\ %[0.5ex] 
		\hline\hline
		\tiny D1 & \tiny 02:26:00.00 &\tiny -04:30:00.0 &\tiny 0.027 & \tiny  Deep, VIMOS, SWIRE, GALEX &\tiny 33 &\tiny 33 &\tiny 66 &\tiny 132 &\tiny 66 \\ 
		\tiny D2 &\tiny 10:00:28.60 &\tiny +02:12:21.0 &\tiny 0.018 &\tiny Cosmos/ACS, VIMOS, SIRTF, XMM &\tiny 33 &\tiny 33 &\tiny 66 &\tiny 132 &\tiny 66\\
		\tiny D3 &\tiny 14:19:28.01 &\tiny +52:40:41.0 &\tiny 0.010 &\tiny Groth strip, Deep2, ACS &\tiny 33 &\tiny 33 &\tiny 66 &\tiny 132 &\tiny 66\\
		\tiny D4 &\tiny 22:15:31.67 &\tiny -17:44:05.0  &\tiny 0.027 &\tiny XMM Deep &\tiny 33 &\tiny 33 &\tiny 66 &\tiny 132 &\tiny 66\\
		\hline
	\end{tabular}
	\caption{Summary of the four Deep fields, the last 5 columns show the amount of time in hours being used per pass band (eg. $u,g,r,i,z$) in each field. Such long allotted hours per band ensures extremely deep stacks of images.}
	\label{tab1}
\end{table}

For the angular separation angle between any two SNe Ia pairs we have used the Haversine formula. Let us explain this briefly. Let two supernovae be in the direction $\mathbf{n_1}=(b_1, l_1)$ and $\mathbf{n_2}=(b_2, l_2)$, where $b$ and $l$ are latitude and longitude respectively of the supernova in the galactic co-ordinate system. The separation angle between those two supernovae on the surface of the sphere is given by the Haversine formula which determines the great-circle distance between two points on the sphere and is given by
\bea
\cos(\theta_{12})=\sin(b_1)\sin(b_2)+\cos(b_1)\cos(b_2)\cos(l_1-l_2),\ \mbox{where} \ \theta_{12}=|\mathbf{n_1}-\mathbf{n_2}|.
\eea   
%In what follows we shall make use of this data in the  analysis of reconstruction of convergence power spectrum for SNLS weak lensing.
%%%%%%%%%%%%%%%%%%%%%%%%%%%%%%%%%%%%%%%%%%%%%%%%%%%%%%%%%%%%%%%%%%%%%%%%%%%%%%%%%%%%%%%%%%%%%%%
%%%%%%%%%%%%%%%%%%%%%%%%%%%%%%%%%%%%%%%%%%%%%%%%%%%%%%%%%%%%%%%%%%%%%%%%%%%%%%%%%%%%%%%%%%%%%%

\section{Analysis with original dataset}\label{methodology}
In this section we shall follow the technique developed in Refs. \cite{kaiser1992,schneider2002} and the references therein. Though  their analysis mostly deals with two point statistics of cosmic shear, we have employed to some extent, a  similar method to determine the two point statistics of the lensing convergence field, which is our point of consideration. The dataset we choose in order to materialize this
consists of the real data points for 5-year SNLS data as discussed in the previous section. Using this we shall show how one can extract out meaningful information for two point lensing correlation function.  This will also help us in comparing the results with the analysis done
in the rest of the article using spatial resampling and see if we can improve on the results as obtained from the original dataset.
We will elaborate on this and will justify doing so in due course. 

%%%%%%%%%%%%%%%%%%%%%%%%%%%%%%%%%%%%%%%%%%%%%%%%%%%%%%%%%%%%%%%%%%%%%%%%%%%%%%%%%%%5
\subsection{Two-point convergence correlation function}
The  two-point angular correlation function for the convergence of two sources located at ${\bm \vartheta}$ and $\bm{ \vartheta+\theta}$, can be defined as \cite{kaiser1992}
\bea
\xi_\kappa(\theta)\equiv\langle\kappa(\bm\vartheta)\kappa(\bm{ \vartheta+\theta})\rangle, 
\eea
where $ \ \theta=|\bm{\theta}|$ is the angular separation between the two sources under consideration.
For the power spectrum we transform the  convergence field in a 2D Fourier transform as follows,
\bea
\kappa_\ell=\int d^2(\bm\theta) \kappa(\bm\theta)\exp(-i{\bm \ell \cdot \bm\theta})
\eea
where $\ell$ is a 2D dimensionless wavenumber. The power spectrum of the convergence field, $P_\kappa$, can then be found from the following equation
\bea
\langle\kappa_\ell\kappa^{*}_{\ell'}\rangle =\left(2\pi\right)^2\delta\left({\bm\ell-\bm\ell'}\right)P_\kappa(\ell) 
\eea
With the help of the above equation, it is now very easy to see that the correlation function is related to the convergence power spectrum through the following relation
\bea
\xi_\kappa(\theta)=\frac{1}{2\pi}\int_{0}^{\infty}\ell d\ell {P}_\kappa(\ell)J_0(\ell \theta).
\eea
The above relation can be inverted using the orthogonal property of the Bessel function and the corresponding expression is as follows:
\bea
{P}_\kappa(\ell)=2\pi\int_{0}^{\infty}\theta d\theta \xi_\kappa(\theta)J_0(\ell \theta)
\eea
This provides a straightforward route  to estimate the power spectrum directly from the observable correlation function. However,  one needs to have the knowledge of the correlation function at all angles.  

In what follows we shall make use of the above methodology to extract out meaningful information on the two point statistic for Type Ia supernovae lensing convergence. In the absence of lensing $B$-mode all the second order lensing statistics are interrelated and they can be estimated from the convergence alone \cite{schneider2002}.  Also, the shear due to gravitational lensing of the large scale structure is pure gradient field, $E$-mode, therefore no $B$-modes should be there. Hence, our assumption of no $B$-mode in the weak lensing limit is very nigh to the real fact. 

%%%%%%%%%%%%%%%%%%%%%%%%%%%%%%%%%%%%%%%%%%%%%%%%%%%%%%%

\subsection{Estimator}\label{estimator}
Assuming that the correlation function is to be estimated in bins of angular width $\delta\theta$, the estimator for the convergence correlation function can be defined as \cite{schneider2002}
\bea
\hat{\xi}_\kappa(\theta)&=&\frac{\sum_{i>j}\kappa(\bm{\vartheta_i})\kappa(\bm{\vartheta_j})\Delta_\theta(|\bm{\vartheta_i-\vartheta_j}|)}{N_\kappa(\theta)}, \ \mbox{where}\label{estimator_kappa}\\
\Delta_\theta(\phi)&=& 1, \ \mbox{for} \ \theta-\delta\theta/2 <\phi\leq \theta+ \delta\theta/2, \ \mbox{and}  \ 0 \ \mbox{otherwise}; \nonumber \\ &&\mbox{and} \ N_\kappa(\theta)=\sum_{i>j}\Delta_\theta(|\bm{\vartheta_i-\vartheta_j}|).\nonumber
\eea
$N_\kappa(\theta)$ represents number of supernovae pairs in the bin with center at $\theta.$ 
The above sum is performed over all supernova pairs $(i,j)$ with angular separation $|\bm{\vartheta_i-\vartheta_j}|$ within a bin width of $\delta\theta$ around $\theta$. 
\iffalse
In a similar way the estimator for the shear correlations `$\pm$' can be defined 
\bea
\hat{\xi}_\pm(\theta)&=&\frac{\sum_{i>j}\left(\epsilon_t(\bm{\vartheta_i})\epsilon_t(\bm{\vartheta_j})\pm \epsilon_\times(\bm{\vartheta_i})\epsilon_\times(\bm{\vartheta_j})\right)\Delta_\theta(|\bm{\vartheta_i-\vartheta_j}|)}{N_\kappa(\theta)}
\eea
where $\epsilon_{t,\times}$ are tangential and cross-components of the shear. 
Since we shall be using magnification  of the supernovae, we do not directly calculate the `$\pm$' correlations. Instead, we estimate convergence correlation from the data using \eq{estimator_kappa} and then, in principle, `$-$' correlations can be derived from \eq{xm_xp} using the fact that $\xi_\kappa=\xi_+$ in the absence of lensing $B$-mode. But we do not derive the shear correlation functions in this article as our primary intention is to estimate two point correlation function of the convergence itself.
\fi
The advantage of working with the real space correlation function is that, it can be calculated directly from the data.
%%%%%%%%%%%%%%%%%%%%%%%%%%%%%%%%%%%%%%%%%%%%%%%%%%%%%%%%%%%%%%%%%%%%%%%%%%%%%%%%%%%%%%%%%%%%
%%%%%%%%%%%%%%%%%%%%%%%%%%%%%%%%%%%%%%%%%%%%%%%%%%%%%%%%%%%%%%%%%%%%%%%%%%%%%%%%%%%%%%%%%%%%
\subsection{Results}\label{analysis}
In this section, we analyze the data in a different approach as our data set is small containing only $296$ supernovae. We apply internal covariance estimator \cite{friedrich2015} for lensing convergence  without spatial re-sampling. As a result, the error on two point correlation function of the lensing convergence that we report here is only the noise components of the covariance which is equivalent to the shape noise for any cosmic shear survey.

%%%%%%%%%%%%%%%%%%%%%%%%%%%%%%%%%%%%%%%%%%%%%%%%%%%%%%%%%%%%%%%%%%%%%%%%%%%%%%%%%%%%%%%%%%%%%%%%%%% %%%%%%%%%%%%%%%%%%%%%%
%\subsection{The \texorpdfstring{$\xi_\kappa$}{Lg} Correlation and the Error}
To have the estimate of standard error, or the error of the mean of ${\xi}_
\kappa(\theta)$, we first estimate the standard deviation from the data in each bin and then divide by the number of supernovae pairs in that bin as follows 
\bea
\label{e20}
\bar{\xi}_\kappa(\theta_i)&\equiv& \frac{1}{n_{b_i}}\quad\quad\ \ \ \sum_{\mbox{\scriptsize pairs $m, n$ with separation} \ \theta_i}  \kappa_m\kappa_n \\
\sigma^2_{\bar{\xi}_\kappa}(\theta_i)&\equiv& \frac{1}{n_{b_i}-1}\quad\sum_{\mbox{\scriptsize pairs $m, n$ with separation} \ \theta_i}\left( \kappa_m\kappa_n-\bar{\xi}_\kappa(\theta_i)\right)^2\\
s^2_{\bar{\xi}_\kappa}(\theta_i)&\equiv& \frac{1}{n_{b_i}}\sigma^2_{\bar{\xi}_\kappa}(\theta_i)
\eea
where $\bar{\xi}_\kappa(\theta_i)$, $\sigma^2_{\bar{\xi}_\kappa}(\theta_i)$ and $s^2_{\bar{\xi}_\kappa}(\theta_i)$  are the mean, variance and square of the standard error respectively of the ${\xi}_\kappa$ correlation in the $i^{\rm th}$ bin  and ${n_{b_i}}$ is the number of supernovae pairs in that bin. 
\begin{figure}
	\begin{center}
		\includegraphics[width=.6\textwidth]{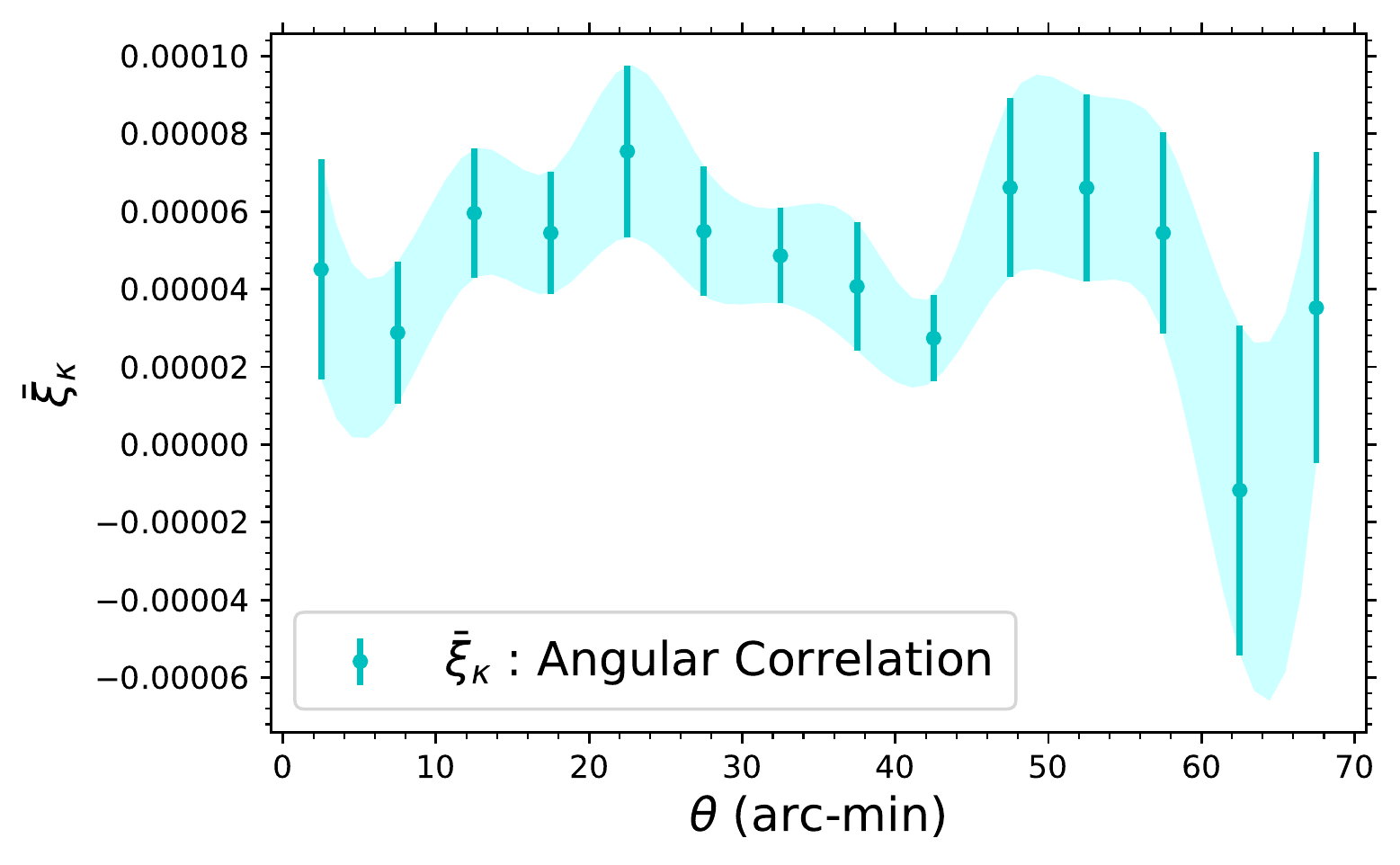}
	\end{center}
	\caption{ We have plotted $\xi_\kappa$ correlations along with its error of the mean and  $\pm1\sigma$ contour. Each bin has a width of $5'$. The plot is based on the original data consisting of 296 supernovae. }
	\label{pm_error}
\end{figure}
In \fig{pm_error} we presented the mean of $\xi_\kappa$ correlation in each bin and the standard error or the error of the mean associated with them along with $\pm1\sigma$ contour. In order to get the mean of $\xi_\kappa$ in each bin, we first computed the $\kappa$ products between every pairs in each field separately (this ensures that we don't compute the angular distance between any two candidates from two different fields),  we combined all the field's $\kappa$ products and the corresponding angular separation  and then performed linear binning based on the angular separation. 

It is apparent from  the \fig{pm_error}  that the convergence field has non-zero value for the 2-pt correlation function at all separation angle which is not consistent with zero. This essentially means that the supernovae convergence  in different angular bins has a definite correlation. In order to make sure that such non-zero correlation indeed  exists we apply spatial resampling technique in the following section. Subsquently, we shall compare between the two results thus obtained and check if we can improve on the original dataset to arrive at our conclusion.
%%%%%%%%%%%%%%%%%%%%%%%%%%%%%%%%%%%%%%%%%%%%%%%%%%%%%%%%%%%%%%%%%%%%%%%%%%%%%%%%%%%%%%%%%%%%%%%%%%%%%%%%%%%%%%%%%%%%%%%%%%%%%%%%%%%%
%%%%%%%%%%%%%%%%%%%%%%%%%%%%%%%%%%%%%%%%%%%%%%%%%%%%%%%%%%%%%%%%%%%%%%%%%%%%%%%%%%%%%%%%%%%%%%%%%%%%%%%%%%%%%%%%%%%%%%%%%%%%%%%%%%%%%%%%%%%%%%%%%%%%%%%%%%%%%%%%%%%%%%%%%%%%%%%%%%%%%%%%%%%%%
%%%%%%%%%%%%%%%%%%%%%%%%%%%%%%%%%%%%%%%%%%%%%%%%%%%%%%%%%%%%%%%%%%%%%%%%%%%%%%%%%%%%%%%%%%%%%%%%%%

\section{Analysis with Spatial Resampling }\label{bootstrap}
In this section we shall analyse the data from a different perspective, adopting spatial resampling method we compute the correlation function from the multi-fold data, which then will be used to  produce the covariance matrix. There are different ways to get the internal error estimates.
Some of them includev subsample method, jackknife technique etc. In this article we make use of bootstrap resampling that is based on the principle of shuffling the original data $N$ number of times. In principle there is no limit on $N$ for  bootstrap error estimates \cite{norberg2009statistical}. But there is a rough lower bound on the number of reshuffling of the data to be used which depends on the accuracy of the parameter to be determined using bootstrap \cite{taylor2013putting, taylor2014estimating}.  In what follows we shall briefly describe how we have utilised the bootstrap resampling technique  for our dataset under consideration.  We shall also compare among different values of $N$ chosen and check the improvement of results.

%%%%%%%%
\iffalse
In order to compute the covariance matrix from the data, it is evident that the sample size is  too small to use. For addressing this shortage of data, we devised this bootstrap method.   We have $296$ SNe-Ia at our disposal.  We perform the bootstrap reshuffling on the $\kappa$'s in each patch separately keeping the sky position and the redshift of the supernovae fixed, i.e. we  only shuffle the $\kappa$ values. Each shuffling enables us to obtain a new realization of the data. After $N$ such shufflings, for each patch separately, we compute the $\kappa$-products, $\kappa_i\kappa_j$, between every pairs $(i,j)$ within a column (i.e. we do not mix different realizations yet)  and their corresponding angular separations, $\Delta\theta_{ij}$. After this, we compute the sum of the $\kappa$-products along each row (i.e. along each $\Delta\theta_{ij}$) and compute the corresponding mean, $\langle\kappa_i\kappa_j\rangle$, by dividing the sum with $N$, 
\bea
\bar{\xi}_\kappa(\Delta\theta_{ij})=\langle\kappa_i\kappa_j\rangle=\frac{1}{N}\sum_{\alpha=1}^{N}\kappa_{i}^\alpha\kappa_{j}^\alpha.
\eea
We repeat the above procedure for each of the four patches separately. Now we mix them to get bigger set containing the two-point angular correlation function of the convergence and separation angles.  
\fi
%%%%%%

\subsection{Bootstrap Resampling}
In order to compute the covariance matrix from the data, it is evident that the sample size is  too small to use. For addressing this shortage of data, we devised this bootstrap method.   We have $296$ SNe-Ia at our disposal.  We perform the bootstrap reshuffling on the $\kappa$'s in each patch separately keeping the sky position and the redshift of the supernovae fixed, i.e. we  only shuffle the $\kappa$ values. Each shuffling enables us to obtain a new realization of the data. After $N$ such shufflings, for each patch separately, we compute the $\kappa$-products, $\kappa_i\kappa_j$, between every pairs $(i,j)$ within a column (i.e. we do not mix different realizations yet)  and their corresponding angular separations, $\Delta\theta_{ij}$. This completes the shuffling bootstrap method.  This shuffling step was carried out with two principal motivations, to be able to use sufficient data to produce a covariance matrix and  to check for the consistency(and compare) of the estimate obtained from this sample at different $N$.  

After taking the $\kappa$ products for each realizations and each patch separately, we bin them into $n_b$ bins along the angular separation, based on a chosen  bin width. Before binning we mix 4-patches in the following way: each realizations with the same marking of the different patches are merged together i.e. a specific  column of the $\kappa$ products of the patch 1 is merged with the corresponding columns of the patches 2, 3, and 4. So that we have  the following set  $\{\theta_i, \ \xi^1(\theta_i), \xi^2(\theta_i),.... \xi^N(\theta_i); \ i=1,2,.... n_b\}$, where $N$ is the number of bootstrap realizations and $\theta_i$ is the centre of the $i^{th}$ bin. Now we calculate the mean of the correlation function in each bin separately as follows
\bea
\bar{\xi}_i\equiv\bar{\xi}(\theta_i)=\frac{1}{N}\sum_{\alpha=1}^N\xi^\alpha(\theta_i), \ \mbox{where} \ i=1,2, .... n_b. \eea
\subsection{Covariance Matrix}
Once the binning is performed and the corresponding mean correlation has been estimated, the last step involves computation of the covariance between the different bins.
The covariance matrix between the $i^{th}$ and $j^{th}$ bins for N bootstrap resamplings  is then defined as:
\bea
C_{ij} = \frac{1}{N-1}\sum_{\alpha=1}^{N}(\xi^{\alpha}_i-\bar{\xi_i})(\xi^{\alpha}_j-\bar{\xi_j})
\eea
We emphasize on the fact that this step to shuffle the data and produce new realizations, are in no way done with the motivation to substitute the lack of original data. But this bootstrapping is done  with the idea to populate the estimator statistics to be able to produce a meaningful covariance matrix from it at the same  signal level. This also enables us to get a cleaner inference from the data about the estimator behaviour for the given signal level.  We will see in the following section how changing the value of $N$ affects the covariance matrix. Further, in the present analysis we have not taken into consideration  any possible variations due to redshift dependence along the line of sight. However future analysis  taking into consideration of the redshift variability, could show results with more accuracy. 

\subsection{Results}
\begin{figure}
     \centering
     \begin{subfigure}[t]{0.48\textwidth}
         \centering
         \includegraphics[width=\textwidth]{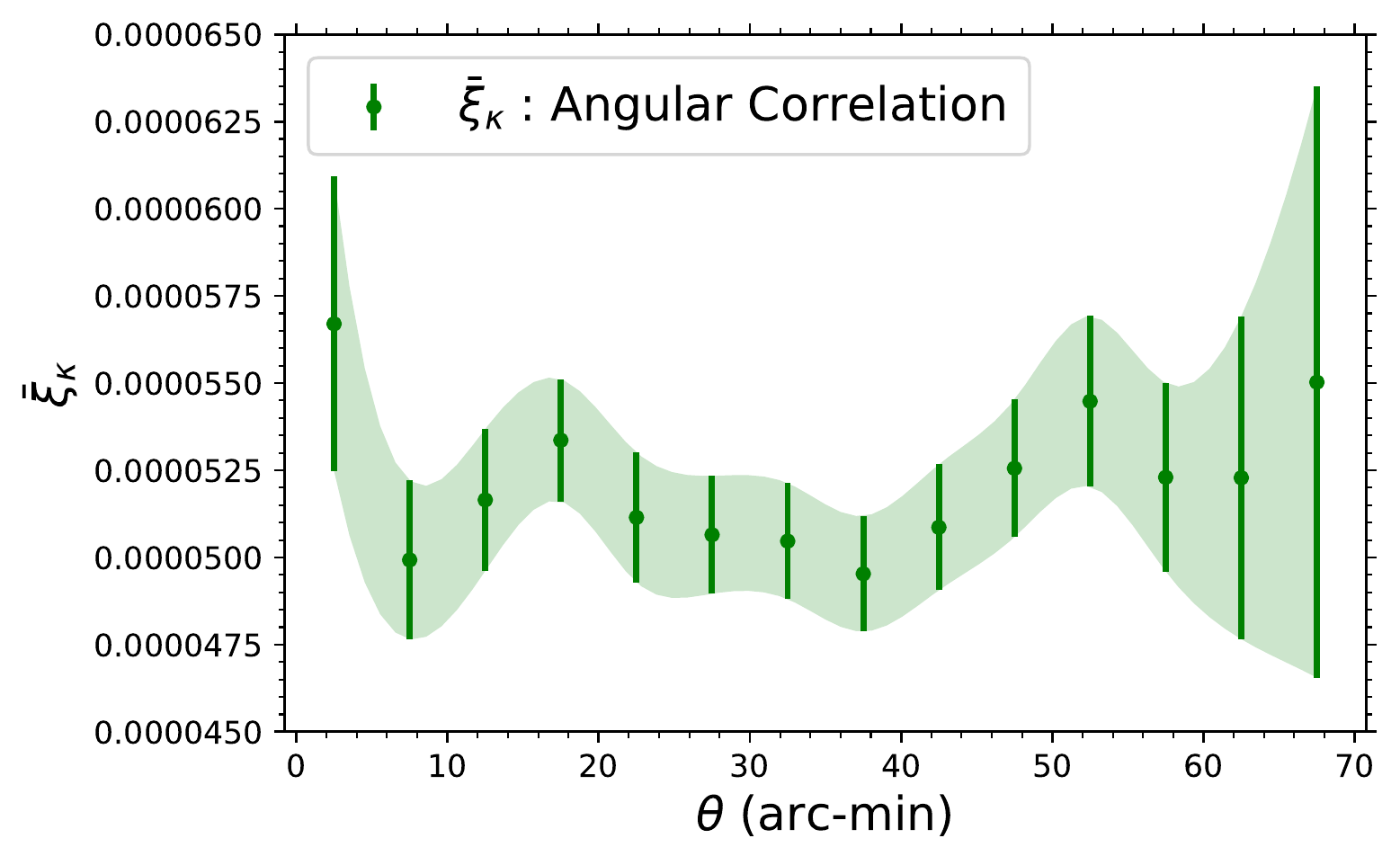}
         \caption{Convergence correlation function estimator for bootstrap reshuffling parameter $N=10^2$ along with $\pm1\sigma$ contour.}
         \label{}
     \end{subfigure}
     \hfill
     \begin{subfigure}[t]{0.48\textwidth}
         \centering
         \includegraphics[width=\textwidth]{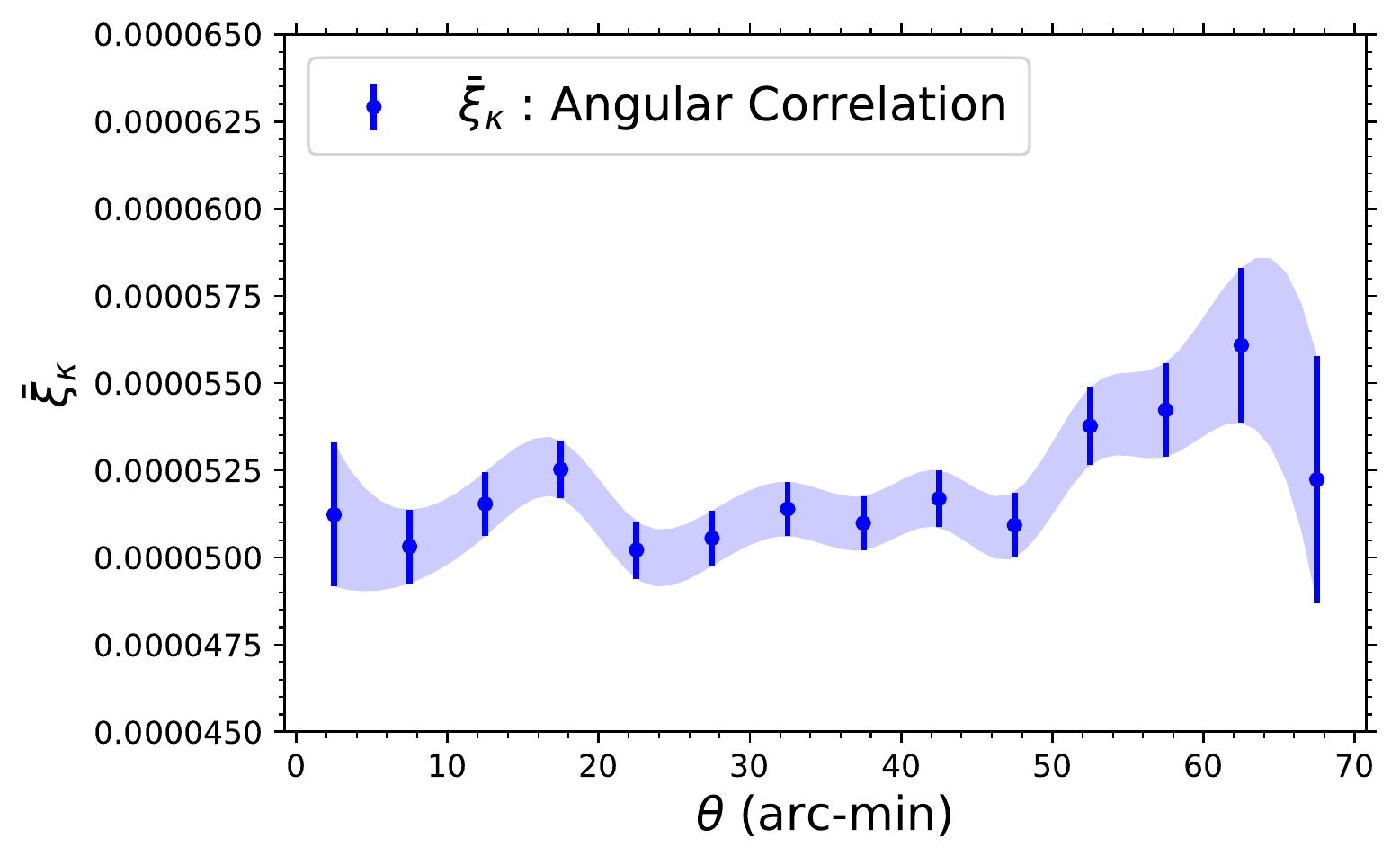}
         \caption{Convergence correlation function estimator for bootstrap reshuffling parameter $N=10^3$ along with $\pm1\sigma$ contour.}
         \label{}
     \end{subfigure}
     \hfill
     \begin{subfigure}[t]{0.48\textwidth}
     \centering
         \includegraphics[width=\textwidth]{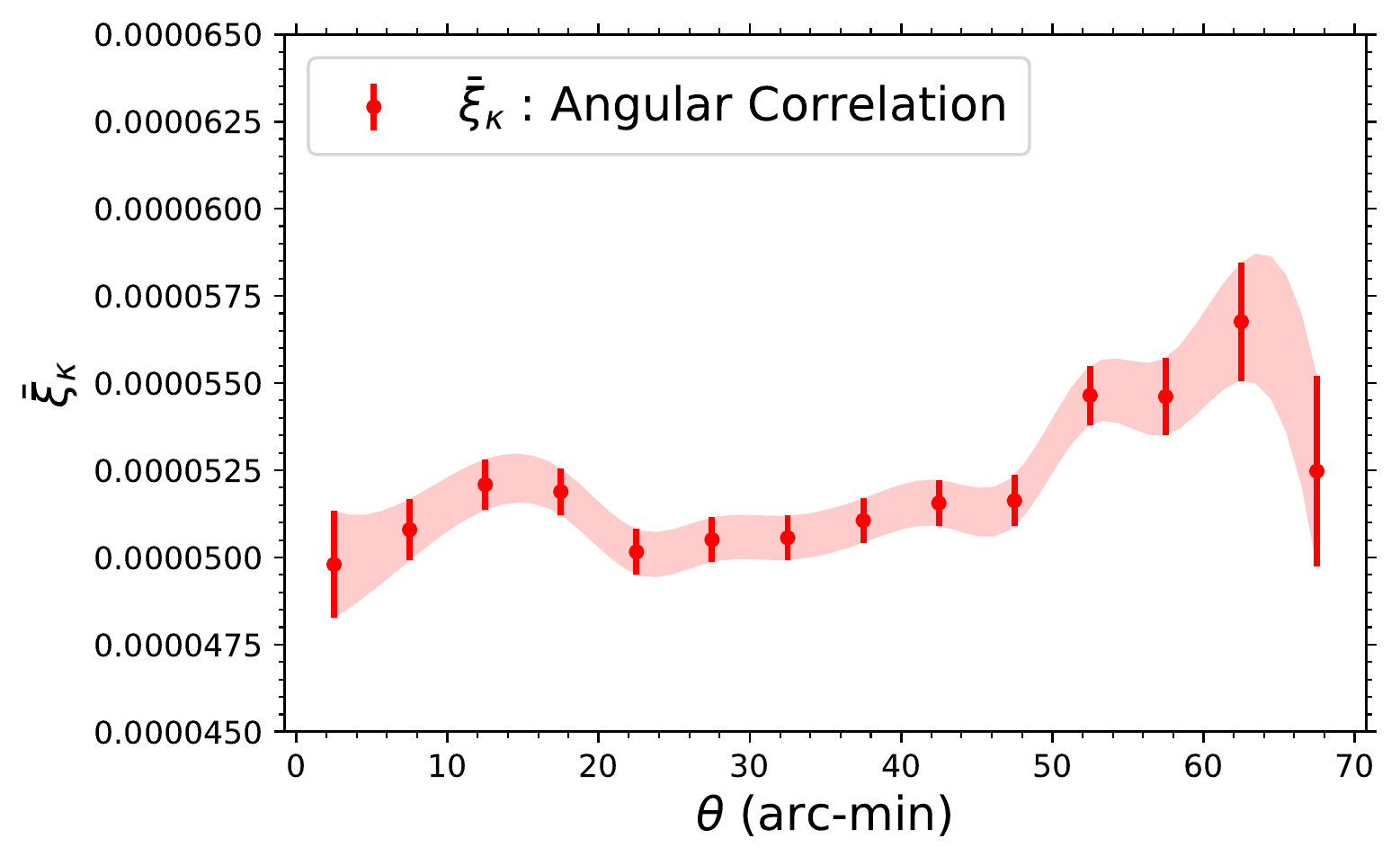}
         \caption{Convergence correlation function estimator for bootstrap reshuffling parameter $N=10^4$ along with $\pm1\sigma$ contour.}
         \label{}
     \end{subfigure}
     \hfill
     \begin{subfigure}[t]{.48\textwidth}
  \centering
   \includegraphics[width=\textwidth]{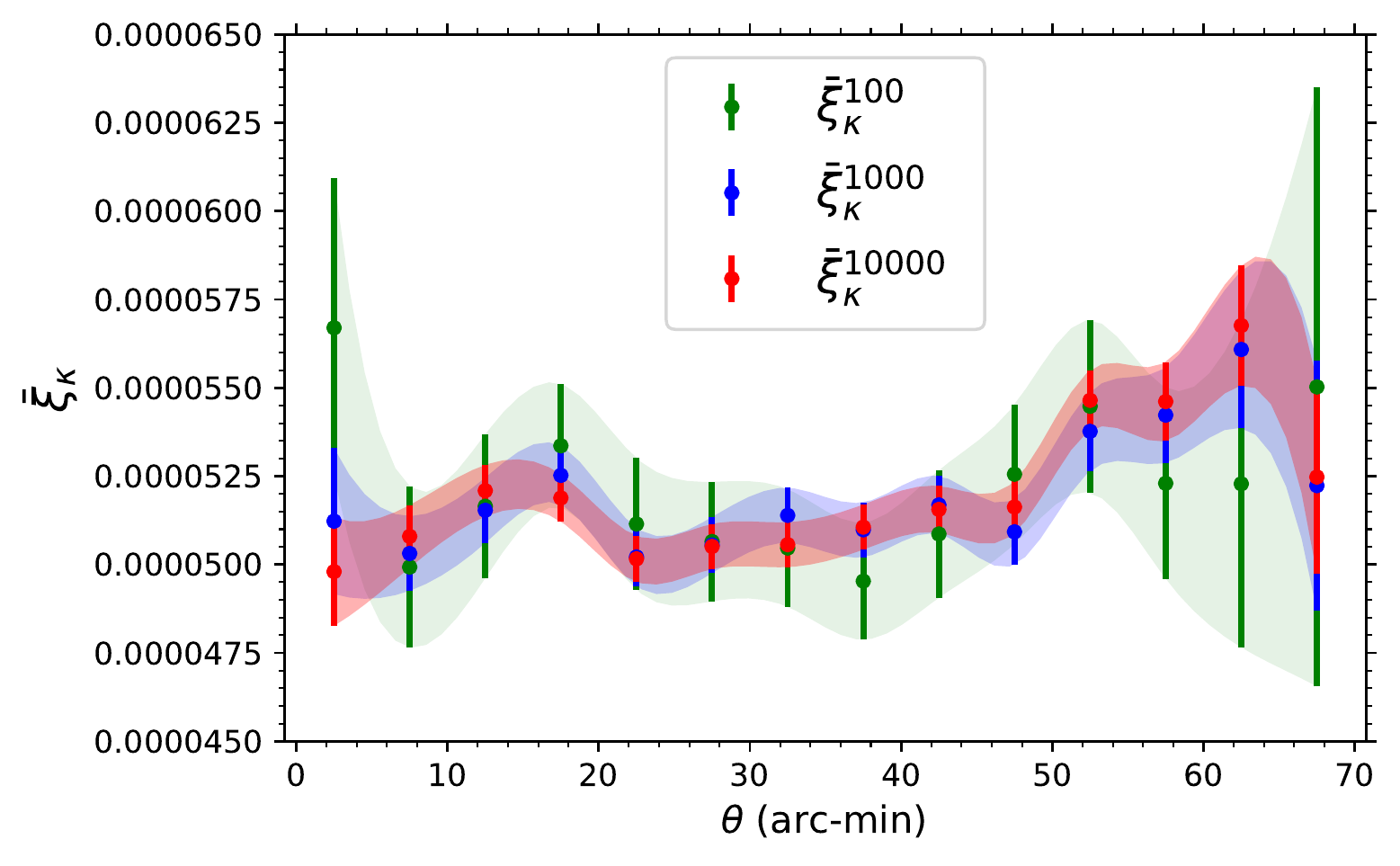}  
  \caption{Comparison of the convergence correlation function estimator when the data is shuffled for $N={10^x},x=[2,3,4]$.   We see that for the $x=[3,4]$ cases the error bars gets shortened(as expected), however they lie within the error margins of the estimator value for the case $x=2$.}
  \label{comparison}
\end{subfigure}
        \caption{
        Plots showing the convergence correlation function estimators for logarithmically increasing values of the bootstrap reshuffling parameter $N$ from $10^2$ to $10^4$ along with $\pm1\sigma$ contour. Each bin has a width of $5'$. Note that the error bars gets shortened when the original data is shuffled more number of times. The difference of the y-axis in the three plots are also to be carefully observed.
        }
        \label{correlations}
\end{figure}

The \fig{correlations} shows the variation of the convergence auto-correlation function with the angular separation  along with $\pm1\sigma$ contour for logarithmically increasing values of the bootstrap reshuffling parameter $N$ from $10^2$ to $10^4$. For our analysis we have applied linear binning with a bin width of $5'$. The plot also reveals that the increasing the number of realizations shrinks the error bar on the correlation function which is in tune with the expectation as increasing sample numbers decreases the spread. In \fig{comparison} we have compared the results for $N=100,1000, 10000$ respectively. Oe can readily check from the figure that the 2-point convergence correlation of the Type Ia supernovae does not vanish irrespective of the number of shuffling. 

%%%%%%%%%%%%%%%%%%%%%%%%%%%%%%%%%%%%%%%%%%%%%%%%%%%%%%%%%%%%%%%%%%%%%%%%%%%%%%%%%%%%%%%%%%%%%%%%%%%%%%%%%%%%%%%%%%%
\section{Discussions and Comparison}\label{com}
\begin{figure}%[h!]
\centering
%\begin{subfigure}[t]{.48\textwidth}
  %\centering
  % include first image
  \includegraphics[width=.6\textwidth]{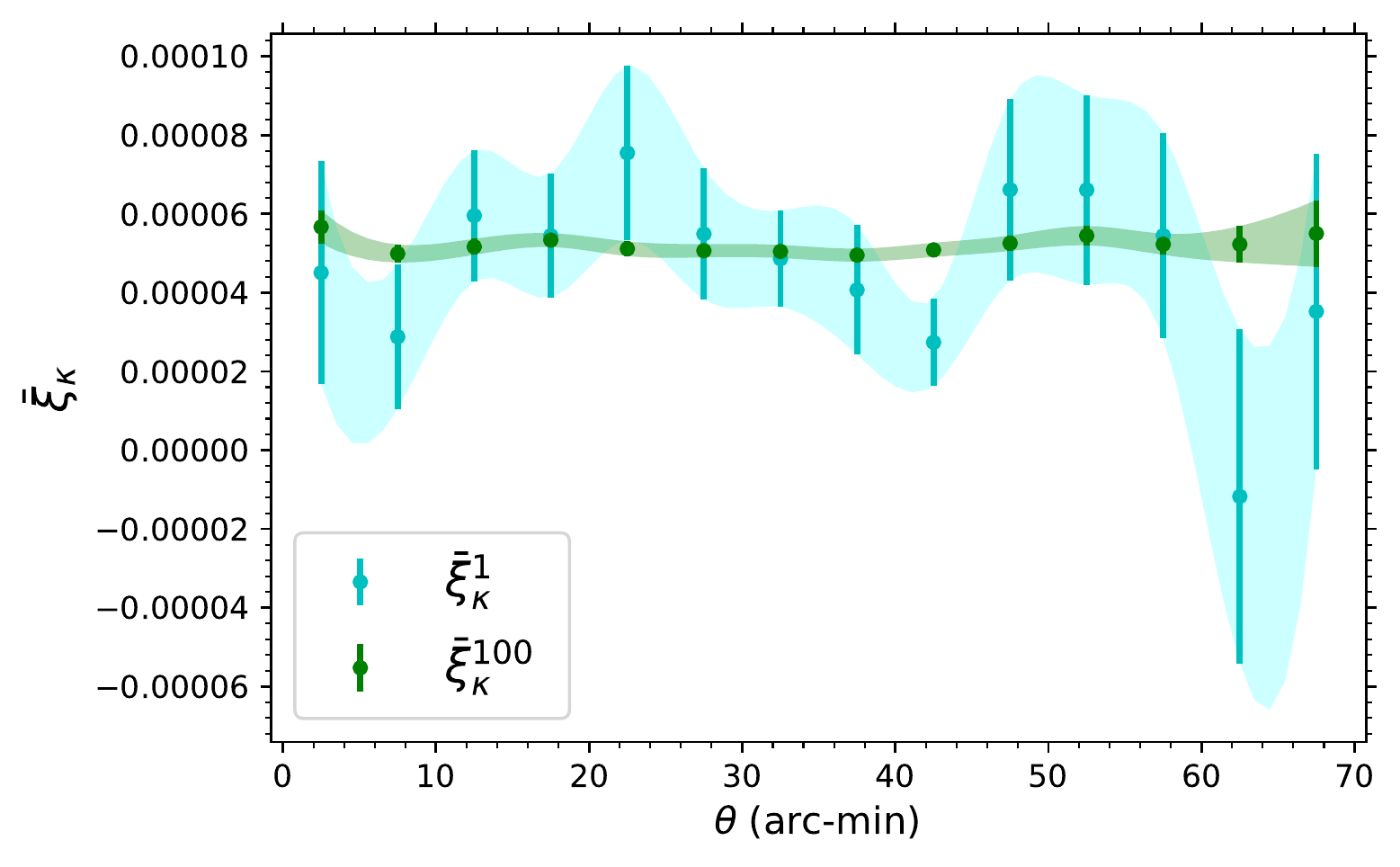}
  \caption{Comparison of the convergence correlation function estimator when the data is shuffled for $N={10^x},x=[0,2]$. The case for $x=0$ is the original scenario, without any shuffling of the data (\fig{pm_error}). It is interesting to note that in the higher angular bins, the estimator is scattered with larger error bins for $x=0$ case. Increasing $x$ makes the behaviour around the higher bins more consistent as we can see in the next plot where we compared $x=2,3,4$.}
\label{cor_compare}
\end{figure}

The angular correlation functions with and without spatial re-samplings are compared in  \fig{cor_compare} as a function of the separation angle. For the comparison we have used 100 bootstrap realizations.  We see that as we go to higher $N$, the error bars ($\sigma$) gets diminished expectedly. We also observe that there is a definite trend in the power distribution. The fact that the power does not vanish at any scale for any value of $N$, gives confidence in the fact that there is a definite presence of signal. The comparative plots (Fig.\ref{correlations}) show us that the choice of the number for $N$ does not influence the signal (fluctuation or vanishing) significantly in magnitude  over the range of $N=[1,10^4]$.  This  implies that one does not necessarily need to go to very high value of $N$ to come to a concrete conclusion or 
to have significant improvement on errors as such. Considering  any value between $N=10^{3}-10^4$ is good enough to arrive at a definite conclusion.

\begin{figure}
     \centering
     \begin{subfigure}[b]{0.32\textwidth}
         \centering
         \includegraphics[width=\textwidth]{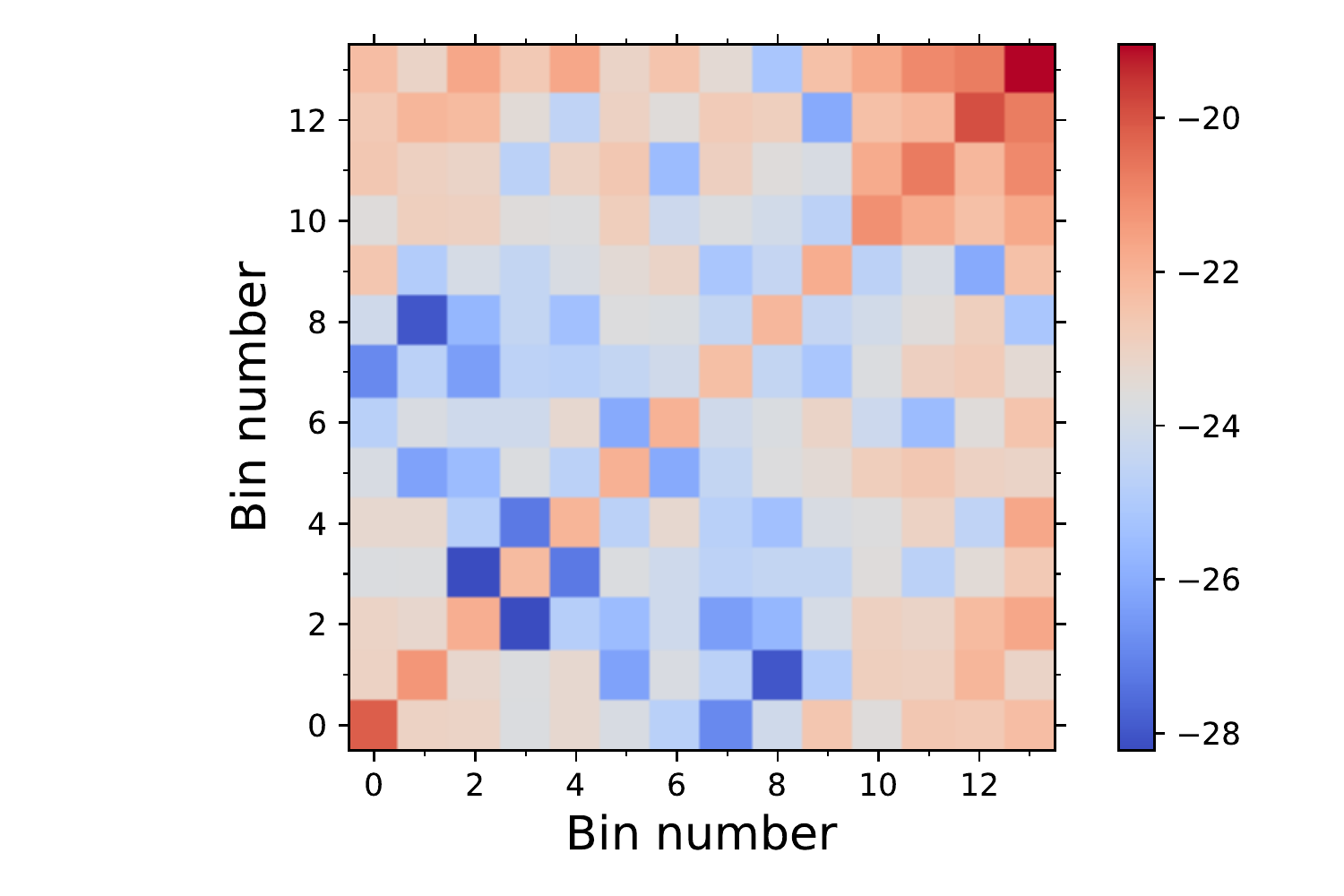}
         \caption{Matrix for visualising the covariance ($C_{ij}$) between the different bins for  bootstrap reshuffling parameter $N=10^2$.}
         \label{}
     \end{subfigure}
     \hfill
     \begin{subfigure}[b]{0.32\textwidth}
         \centering
         \includegraphics[width=\textwidth]{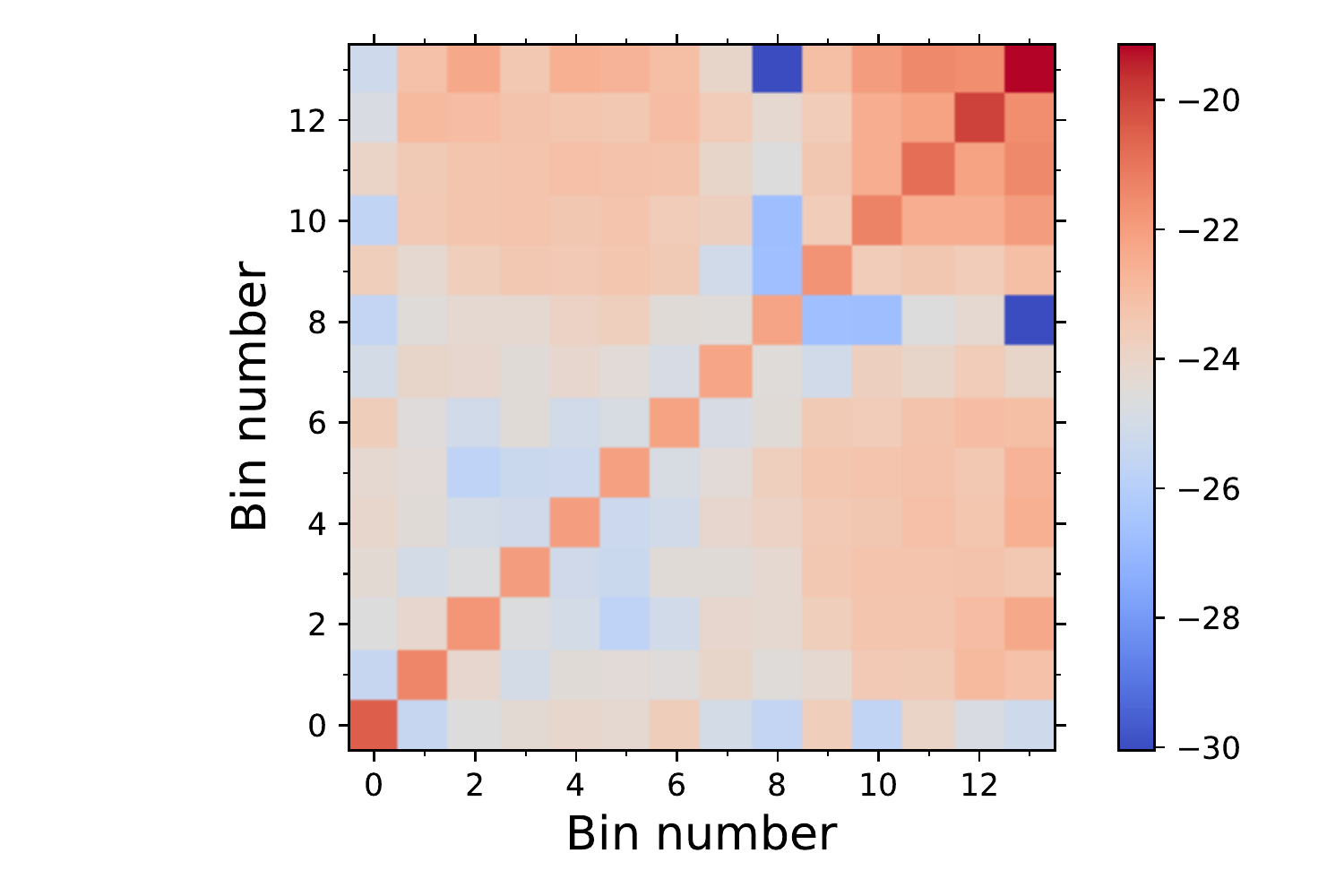}
         \caption{Matrix for visualising the covariance ($C_{ij}$) between the different bins for  bootstrap reshuffling parameter $N=10^3$.}
         \label{}
     \end{subfigure}
     \hfill
     \begin{subfigure}[b]{0.32\textwidth}
         \centering
         \includegraphics[width=\textwidth]{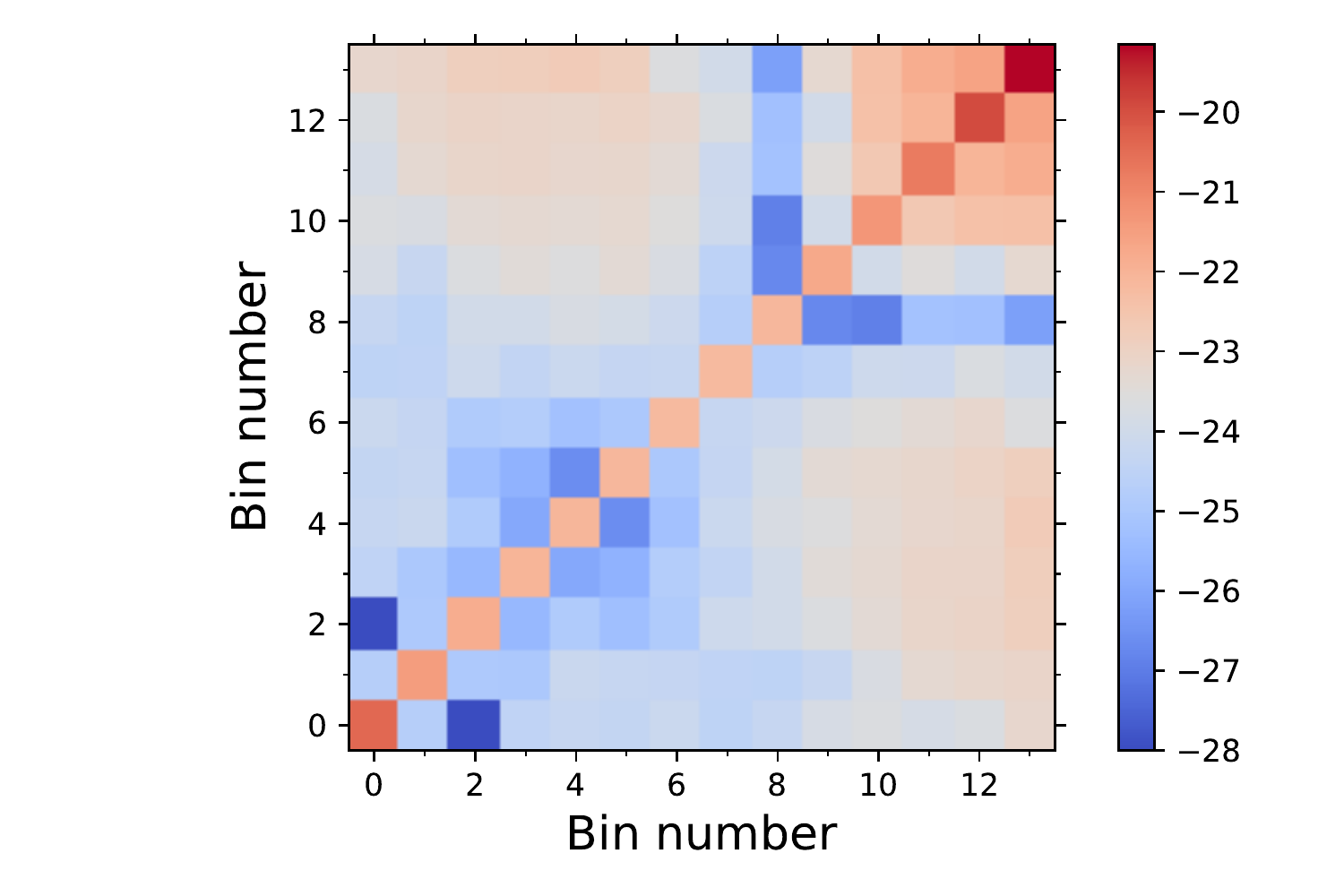}
         \caption{Matrix for visualising the covariance ($C_{ij}$) between the different bins for  bootstrap reshuffling parameter $N=10^4$.}
         \label{}
     \end{subfigure}
        \caption{Matrices for visualising the covariance ($C_{ij}$) between the different bins for logarithmically increasing values of the bootstrap reshuffling parameter $N$ from $10^2$ to $10^4$. The colours of the elements of the matrices
 represent the logarithm of absolute value of the covariances $C_{ij}$. It is interesting to note that the contrast between the diagonal elements increase with increment of the bootstrap reshuffling parameter $N$.}
        \label{covariance}
\end{figure}

In \fig{covariance} we have plotted the covariance matrices for different values of $N$. The covariance matrices have definite diagonal domination. 
This is in tune with the fact that the auto-correlation of the signal should give rise to maximum strength and thus validates our methodology.  
Further, it is interesting to note that the contrast between the diagonal elements increase with increment of the bootstrap reshuffling parameter $N$. This also justifies the use of bootstrap in improving on the results.

\begin{figure}
	\begin{center}
	\includegraphics[width=.6\textwidth]{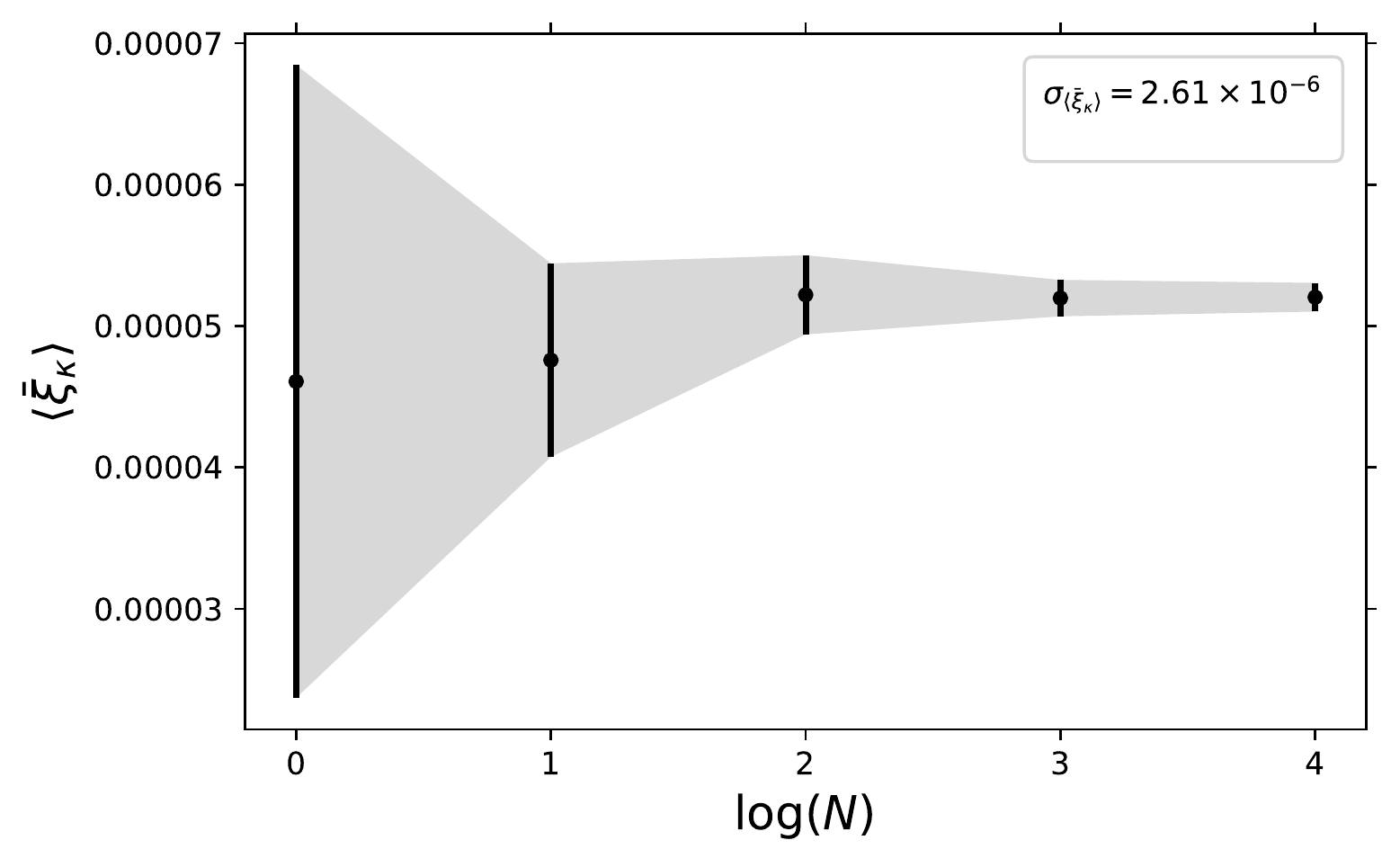}
	\end{center}
	\caption{ We have plotted $\langle\xi_\kappa\rangle$ along with its mean error as a function of $\log(N)$. Roughly it can be seen, that the variation in the estimator (mean $\xi$)  over the values of $N$ is $\sim\mathcal{O}(10^{-1})$.
	}
	\label{error}
\end{figure}

In \fig{error} we have plotted the $\pm1\sigma$ contour for different values of the resampling parameter $N$. From the figure it is clear that as we increase the sample size we get narrower contour. It is also interesting to note that the standard deviation of the five $\langle \bar{\xi}_\kappa\rangle$ values, namely $\sigma_{\langle \bar{\xi}_\kappa \rangle}$ is one order of magnitude smaller than the mean $\langle \bar{\xi}_\kappa\rangle$. This means that the convergence correlation function estimator does not deviate significantly as we increase $N$.
As already pointed out, this  implies that considering  any value between $N=10^{3}-10^4$ is good enough to arrive at a definite conclusion,
as has been done in the present article.

It is now straightforward to compare the entire result as obtained from spatial resampling technique using 296 SN-Ia data points with the  results obtained from the original dataset (without using bootstrap)  in Section \ref{methodology}. First of all, 
 we could construct the covariance matrix  for SNLS 5-year weak lensing data even with the original 296 data points
  using bootstrap, and find proper features of  covariance matrix (e.g., the diagonal terms are dominating) as expected. As obvious, this was not possible to obtain only with the original dataset without spatial resampling. 
 Secondly, 
we could improve on the errors   to a great extent compared to the original dataset consisting of 296 data points and could justify a possible number of times one needs to do the shuffling of data in order to arrive at a reliable conclusion.

%\clearpage
%\end{adjustbox}

%%%%%%%%%%%%%%%%%%%%%%%%%%%%%%%%%%%%%%%%%%%%%%%%%%%%%%%%%%%%%%%%%%%%%%%%%%%%%%%%%%%%%%%%%%%%%%%%%%%%%%%%%%%%%%%%%%%%%%%%%%%%%%%%%%%%%%%%%%%%%%%%%%%%%%%%%%%%%%%%%%%%%%%%%%%%%%%%%%
\section{Summary and Outlook}\label{conclusion}
In terms of weak lensing analysis, dealing with magnification (hence convergence) is more challenging than 
measuring shear because of the difficulty in computing the small difference in the flux distribution of the source between a lensed (magnified) and an unlensed situation. In this article we have presented a very simple yet useful way to estimate two point correlation function of the lensing convergence from the magnification of SNe Ia using SNLS 5-year data. Our analysis reveals that, supernovae magnification can indeed be used to extract information about the lensing convergence which in turn can help us to get hold of the lensing potential. Thus, the major outcome of the present article is two-fold: using bootstrap we could construct the convergence matrix with proper characteristics for SNLS 5-year weak lensing data as well as could improve on the errors compared to the original dataset consisting of 296 data points.
We reshuffle the original data consisting of 296 data points 100-10000 times and compare the results with that obtained from original data points. We show that this technique helps us arrive at a reliable conclusion on weak lensing convergence even though the original dataset comprises of a small number of data points and paves the way to compute the corresponding covariance matrix with reduced error bars on the correlation function.   
This validates the entire exercise done in the present article and, at the same time, leads to interesting conclusion on weak lensing
convergence correlation from SNLS data.

When applying spatial resampling we did not take into account the redshift information of the supernovae magnification. 
During a more detailed study, the broad redshift range of the SNe sample used here, $z\in[0,1.2]$ must be taken into account. One possible way to deal with this could be by  freezing different snapshots of the redshift slices and analyzing them separately.
However as the first step, our primary intention was to demonstrate that the method indeed helps in extracting out meaningful information about the two point correlation function of the lensing convergence field from SNLS weak lensing data,  also to be able to produce the covariance matrix from the same pipeline.  It is always possible in future to take into account more real world complexities and tighten the expressions presented here.

Our analysis may also help in putting tighter constraints on the cosmological parameters when combined with other datasets, although we did not address that issue in this work.   We wish to come back with  likelihood  analysis of SNLS 5-year data along with its cross-correlation with other data using the redshift information shortly.

As the bottomline, we stress upon the fact  that the analysis presented here could pave the way for future SN surveys where with the abundance of SNe candidates the lensing effect would become one of the major sources of uncertainty. For example the  \cite{lsst} plans to observe SNe in order $\sim10^6$ over the next decade, giving a major boost to the present scenario where we only deal with $\sim300$ SNe Ia. Consequently, with the increasing sample size the accuracy of our formalism is also expected to improve by one, or  more optimistically, a few, order(s) of magnitude at some stage. Which can be seen from the comparison results of the covariance matrices. In future, there will be  availability of data from surveys, corresponding to $N=10^4$ case for constructing a covariance matrix directly from our analysis pipeline.

%\newpage
%%%%%%%%%
\section*{Acknowledgments}
We would like to thank the   \href{http://supernovae.in2p3.fr/}{SNLS Paris group}  from the LPNHE, Sorbonne University, for allowing us to re-use the SNLS data. We are specially grateful to  Dr. Marc Betoule and Prof. Delphine Hardin for their cooperation.  AM carried out this project with  the grant from RK MES grant AP05135753, Kazakhstan.  AM would also like to thank  Prof. Tuhin Ghosh of NISER, Orissa and Prof. Eric Linder for their useful suggestions.  BKP would like to thank IUCAA, Pune for giving him the opportunity  to carry on research work through their Associateship Program. AP is supported by Council of Scientific and Industrial Research (CSIR), India (File no.  09/093(0169)/2015 EMR-I).  

%This research did not receive any specific grant from funding agencies in the public, commercial, or not-for-profit sectors.



%%%%%%%%%%%%%%%%%%%% REFERENCES %%%%%%%%%%%%%%%%%%%%%%%%%%%%

%\newpage
\bibliographystyle{mnras}
\bibliography{references} % if your bibtex file is 

%%%%%%%%%%%%%%%%%%%%%%%%%%%%%%%%%%%%%%%%%%%%%%%%%%

% Don't change these lines
\bsp	% typesetting comment
\label{lastpage}
\end{document}